%% file: NonrelFluid.tex
\documentclass[12pt]{article}

\usepackage{epsfig,graphicx,bm,amsmath,amsfonts}
\usepackage{epstopdf}

\usepackage{cite}
\usepackage{psfrag}
\usepackage{graphicx}
\usepackage{xcolor}
\usepackage{bm}
\usepackage{amssymb}
\usepackage{fancyhdr}
\usepackage{url}
\usepackage{epstopdf}
\usepackage[utf8]{inputenc}
\usepackage{verbatim}

\usepackage{float}
\usepackage[font=small,labelfont=bf]{caption}
\usepackage{mathrsfs}
\usepackage[parfill]{parskip}
\usepackage{subcaption}

\numberwithin{equation}{section}

\input{gdefn.tex}

\begin{document}
	
	\setlength{\unitlength}{1mm}
	
	\begin{titlepage}
		
		\begin{flushright}
			IISER Bhopal\\
			HRI
		\end{flushright}
		\vspace{2cm}
		
		\begin{center}
			{\bf \Large Light-Cone Reduction vs. TsT transformations : A Fluid Dynamics Perspective}\\
			\vspace*{1cm}
		\end{center}

		\begin{center}
			\bf{Suvankar Dutta$^{(a)}$\footnote{suvankar@iiserb.ac.in},}
			\bf{Hare Krishna$^{(b)}$\footnote{harekrishna@hri.res.in}}
			
			\vspace{.5cm}
			
			{\small \it $^a$Indian Institute of Science Education and Research Bhopal\\
				Bhopal Bypass, Bhauri,
				Bhopal 462066\\
			\vspace{2mm}
			{\small \it $^b$Harish-Chandra Research Institute\\
				Chhatnag road, Jhunsi, 
				Allahabad 211019}\\
			}			
		\end{center}
		
		\vspace{1cm}
		
		\begin{abstract}
					 We compute constitutive relations for a charged $(2+1)$ dimensional Schr\"odinger fluid up to first order in derivative expansion, using holographic techniques. Starting with a locally boosted, asymptotically $AdS$, $4+1$ dimensional charged black brane geometry, we uplift that to ten dimensions and perform $TsT$ transformations to obtain an effective five dimensional local black brane solution with asymptotically Schr\"odinger isometries. By suitably implementing the holographic techniques,  we compute the constitutive relations for the effective fluid living on the boundary of this space-time and extract first order transport coefficients from these relations. Schr\"odinger fluid can also be obtained by reducing a charged relativistic conformal fluid over light-cone. It turns out that both the approaches result the same system at the end. Fluid obtained by light-cone reduction satisfies a restricted class of thermodynamics. Here, we see that the charged fluid obtained holographically also belongs to the same restricted class.  
		\end{abstract}
		
		\vspace{2cm}
		
		{\bf{\small Keywords:}} Fluid-gravity correspondence, Non-relativistic fluid dynamics
		
	\end{titlepage}

	\tableofcontents

\input{Intro.tex}

\input{lc_rev.tex}

\input{charged.tex}

\input{compare.tex}

\paragraph{Acknowledgement} We are grateful to Arghya Chattopadhyay for collaborating at the initial stage of this project. We would like to thank Nabamita Banerjee, Akash Jain for many helpful discussion. SD acknowledges the Simons Associateship, ICTP. Work of SD is supported by DST under a project with number {\it EMR/2016/006294}. Finally, we are indebted to people of India for their unconditional support towards researches in basic sciences.



\bibliography{bibforTsT}{}
\bibliographystyle{jhep}

\end{document}

%% file: gdefn.tex
\newcommand{\nn}{\nonumber \\}

\newcommand{\ben}{\begin{eqnarray}\displaystyle}
\newcommand{\een}{\end{eqnarray}}

\newcommand{\be}{\begin{equation}}
\newcommand{\ee}{\end{equation}}

\newcommand{\bc}{\begin{center}}
\newcommand{\ec}{\end{center}}

\newcommand{\eesp}{\end{split}}
\newcommand{\bsp}{\begin{split}}

\newcommand{\Rmnum}[1]{\expandafter\@slowromancap\romannumeral #1@}

\renewcommand{\a}{\alpha}	
\renewcommand{\b}{\beta}		
\renewcommand{\d}{\delta}	
\newcommand{\dow}{\partial}
\newcommand{\pa}{\partial}
\newcommand{\e}{\epsilon}

\newcommand{\g}{\gamma}

\renewcommand{\k}{\kappa}	
\renewcommand{\l}{\lambda}	
\newcommand{\m}{\mu}								
\newcommand{\n}{\nu}								
\newcommand{\vp}{\varpi}							

\renewcommand{\r}{\rho}		
\newcommand{\s}{\sigma}

\renewcommand{\t}{\tau}		

\newcommand{\cA}{\mathcal{A}}

\newcommand{\cC}{\mathcal{C}}
\newcommand{\cD}{\mathcal{D}}

\newcommand{\cF}{\mathcal{F}}
\newcommand{\cG}{\mathcal{G}}

\newcommand{\cI}{\mathcal{I}}

\newcommand{\cM}{\mathcal{M}}

\newcommand{\cO}{\mathcal{O}}
\newcommand{\cP}{\mathcal{P}}

\newcommand{\cS}{\mathcal{S}}

\newcommand{\cW}{\mathcal{W}}

\newcommand{\ra}{\rightarrow}

\newcommand{\lB}{\left [}
\newcommand{\rB}{\right ]}
\newcommand{\lb}{\left (}
\newcommand{\rb}{\right )}

\newcommand{\tand}{\text{and}}

\newcommand{\bensp}{\begin{eqnarray}\begin{split}}
\newcommand{\eensp}{\end{eqnarray}\end{split}}

\newcommand{\sch}{ Schr\"odinger\ }

\newcommand{\xp}{x^{+}}
\newcommand{\xm}{x^-}

\def\XXint#1#2#3{{\setbox0=\hbox{$#1{#2#3}{\int}$ }
\vcenter{\hbox{$#2#3$ }}\kern-.6\wd0}}

%% file: Intro.tex
\section{Introduction and summary}\label{sec:intro}

Gauge/gravity duality plays an important role to study different properties of fluid dynamics. The duality has been used extensively to obtain holographic stress-tensor and other conserved quantities up to second order in derivative expansion for a fairly general class of relativistic fluids. Holographic study opens up a new paradigm in fluid dynamics where different parity-odd transports are present. Study of parity-odd hydrodynamics has become a very fascinating topics in recent years. 

Most of the recent developments in fluid dynamics are focused to relativistic systems. Much
attention has not been paid to non-relativistic systems\footnote{Non-relativistic conformal field theories with \sch group symmetry consists of the usual Galilean invariance, the scaling symmetry as well as the particle number symmetry \cite{Duval:1984cj,Duval:2008jg,Duval:2009vt}.}. Non-relativistic fluids, in particular, are interesting as they are expected to be realised in low energy experiments. A non-relativistic systems can be thought of an effective low energy (velocity) description of an underlying relativistic systems. Hence, it is natural to expect that the constitutive relations of a non-relativistic fluid, obtained as an effective description of a relativistic theory, also contain parity-odd terms. There are different ways to take non-relativistic limit of a relativistic system. It is well known that under discrete light-cone reduction (LCR), $(d+1,1)$ dimensional Poincar\'e algebra boils down to $d$ dimensional Galilean algebra. Since hydrodynamics is a low energy fluctuations of equilibrium quantum field theory, LCR of relativistic constitutive relations produces constitutive relations of a Galilean fluid in one lower dimension \cite{Rangamani:2008gi,Brattan:2010bw,Banerjee:2014mka}. All the thermodynamic and hydrodynamic data of non-relativistic fluid are completely fixed in terms of the corresponding relativistic fluid data. However \cite{Banerjee:2014mka} observed that the reduced non-relativistic fluid obtained by LCR follows some restricted class of thermodynamics i.e. thermodynamic variables satisfy an extra equation. This extra condition follows from the demand that reduced non-relativistic fluid satisfies first law of thermodynamics and Euler relation locally. This extra relation implies that LCR provides a particular kind of fluid which satisfies that equation locally. A more general class of non-relativistic fluid\footnote{See also \cite{Geracie:2015xfa}.} can be obtained by light cone reduction if we start with a modified relativistic system called "null fluid" \cite{Banerjee:2015uta,Banerjee:2015hra}. 

Gauge/gravity duality also offers a platform to explore properties of fluids order by order in derivative expansion \cite{Bhattacharyya:2008jc,Banerjee:2008th,Erdmenger:2008rm}. It would be interesting to understand whether holographic technique will result a generic class of Galilean/\sch fluid in comparison to LCR. In this paper we {\it holographically} construct the constitutive relations for a non-relativistic charged fluid up to first order in derivative expansion using the fluid/gravity correspondence. A similar study was done by \cite{Rangamani:2008gi,Brattan:2010bw}, where the authors computed non-relativistic currents and transport coefficients using the {\it LCR dictionary} between relativistic fluid and non-relativistic fluid\footnote{LCR provides a dictionary between non-relativistic fluid variables, transports and those of non-relativistic fluid. See section \ref{sec:review} for details.}. They used holographic values of relativistic transports \cite{Bhattacharyya:2008jc,Banerjee:2008th,Erdmenger:2008rm} to find transports of non-relativistic fluid using the LCR dictionary. A direct computation of fluid constitutive relations from bulk was missing in their work. In this paper, we follow a completely different route. We start with dual holographic geometry of a relativistic charged conformal fluid \cite{Banerjee:2008th,Erdmenger:2008rm}. We uplift this five dimensional solution to ten dimensions to fit as a solution of type IIB string theory. Performing TsT transformations followed by a dimensional reduction over five sphere we obtain an effective five dimensional geometry whose asymptotic symmetry group is not Poincar\'e but Schr\"odinger. This effective five dimensional geometry can be written as a solution of equations of motion obtained from an effective five dimensional action \cite{Herzog:2008wg,Adams:2009dm,Brattan:2010bw}. We find the boundary action for this effective theory for well defined variation. Following \cite{Ross:2009ar}, we compute the constitutive relations for Schr\"odinger fluid from this boundary action . 

In summary, there are two different paths to find constitutive relations for non-relativistic fluid dynamics holographically. 
\begin{itemize} 
	\item Path I : Construct dual holographic geometry of a   relativistic fluid and apply AdS/CFT dictionary to obtain conserved currents and transports of boundary system \cite{Banerjee:2008th,Erdmenger:2008rm}. Obtain non-relativistic currents and transports by using the LCR dictionary between relativistic fluid and \sch fluid \cite{Rangamani:2008gi,Brattan:2010bw}. 
  \item Path II : Consider holographic spacetime for relativistic fluid \cite{Kovtun:2003wp,Son:2006em}. Apply $TsT$ transformation on this geometry to obtain bulk
dual of an asymptotically \sch fluid. Construct conserved current of boundary theory using AdS/CFT prescription \cite{Ross:2009ar} and find out the transports from those currents.
\end{itemize}
\begin{figure}[h]
	\centering
	\includegraphics[width=12cm,height=6cm]{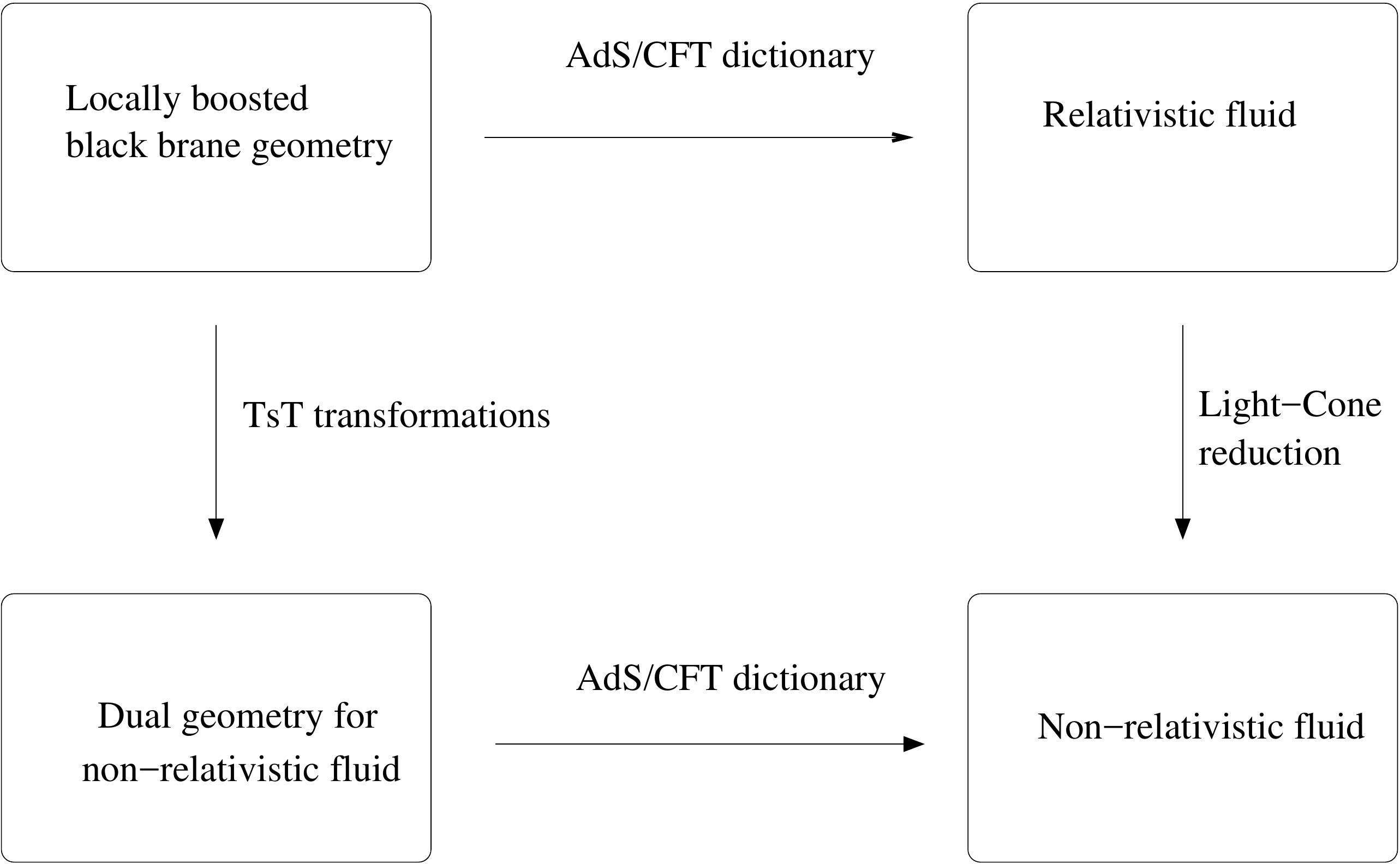}
	\caption{Light-Cone Reduction vs TsT Transformations.}
	\label{fig:dig}
\end{figure}
It would be interesting to understand if these two paths are compatible to all orders in derivative expansion. In this paper, we follow the second path to obtain non-relativistic constitutive relations for a $U(1)$ charged fluid. We observe that path I and path II are same up to first order in derivative expansion. The non-relativistic fluid obtained by LCR has restricted thermodynamics :  LCR of relativistic first law and Euler relation provides correct first law and Euler relation for non-relativistic system at the cost of a constraint relation between non-relativistic energy density, pressure and mass density. In this scenes, LCR provides a restricted thermodynamic system. While obtaining the constitutive relations using holography it turns out that the $'+'$ component of relativistic velocity $u^+(x)$ plays the role of mass chemical potential and terms proportional to derivative of mass chemical potential ($\dow_i u^+$) appear in constitutive relations. We explicitly check that the non-relativistic fluid, obtained holographically is same as that obtained by LCR. We compare our results with \cite{Banerjee:2014mka} and show that the holographic non-relativistic fluid also belong to the same restricted class up to first order in derivative expansion. We summarise our observations in a commutative diagram given in figure \ref{fig:dig}.

Organization of this paper is following. In section \ref{sec:review} we briefly review the main results of LCR of relativistic charged fluid. Construction of bulk geometry for non-relativistic charged fluid has been discussed in section \ref{sec:bulk}. The main results (constitutive relations and transport coefficients for non-relativistic charged fluid) of this paper are given in section \ref{sec:stresstensor}. Finally, we provide a vivid discussion on our results in section \ref{sec:discussion}.

%% file: lc_rev.tex
\section{Light cone reduction of relativistic fluids}\label{sec:review}

In this section we discuss how non-relativistic constitutive relations and transports can be obtained from relativistic data \cite{Rangamani:2008gi,Brattan:2010bw,Banerjee:2014mka}. We start with relativistic constitutive equations and reduce these equations along light-cone direction. Non-relativistic quantities like energy density, pressure etc. are given in terms of relativistic stress tensor.

A conformal relativistic fluid in flat $(d+1,1)$ dimensional spacetime with $U(1)$ isometry is given by the following two conservation equations
\begin{eqnarray}
\begin{split}
\partial_{\mu}T^{\mu\nu}&=0, \\ 
\partial_{\mu} J^{\mu}&=0
\end{split}
\end{eqnarray}
where $T^{\mu\nu}$ is energy momentum tensor, given by
\begin{eqnarray}
\begin{split}
T^{\mu\nu} &= (E+P)u^{\mu}u^{\nu}+P \eta_{\mu\nu}+\Pi^{\mu\nu} 
\end{split}
\end{eqnarray}
and $J^{\mu}$ is conserved $U(1)$ current,
\begin{eqnarray}
\begin{split}
J^{\mu} &= n u^{\mu}+ \Upsilon^{\mu}.
\end{split}
\end{eqnarray}
$E, P$ and $n$ are energy density, pressure and charge density respectively. $\Pi^{\mu\nu}$ and $J^{\mu}$ are dissipative part of energy-momentum tensor and charge current. Up to first order in derivative expansion they are given by,
\begin{eqnarray}
\begin{split}
\Pi^{\mu\nu} &= - \eta_{rel} P^{\mu\alpha}P^{\nu\beta}\left[
\dow_\alpha u_\beta + \dow_\beta u_\alpha -
\frac{2}{d+1} \eta_{\alpha\beta}\dow\cdot u \right], \quad P^{\mu\nu}=\eta^{\mu\nu}+u^\mu u^\nu \\
\Upsilon^\mu & =
-\varrho P^{\mu\nu}\dow_\nu \left(\frac{M}{T}\right)  
- \gamma P^{\mu\nu}\dow_\nu T 
+ \mho l^\mu , \quad l^\mu = \epsilon^{\mu \alpha \beta
	\gamma}u_\alpha\dow\beta u_\gamma
\end{split}
\end{eqnarray}
where $T$ and $M$ are temperature and $U(1)$ chemical potential. $\eta_{rel}, \varrho, \gamma$ are shear viscosity, charge conductivity, thermal conductivity respectively. $\mho$ is a parity-odd transport. Since we are dealing with relativistic conformal fluid (i.e. $T^\mu_{\ \mu}=0$) its bulk viscosity coefficient is zero and the fluid satisfies an equation of state $E=3P$.

We are considering the fluid in a flat background (in absence of any external fields). The metric is given by
\be
ds^2 = -(dx^0)^2 + (dx^{d+1})^2+\sum_{i=1}^{d}(dx^i)^2.
\ee
We introduce the light-cone coordinates,
\be
x^{\pm} = \frac{1}{\sqrt2}\lb x^0 \pm x^{d+1} \rb.
\ee
In the light-cone frame the metric can be written as,
\be
ds^2 = - 2 dx^+dx^-  + \sum_{i=1}^{d} (dx^i)^2 .
\ee
Since, the symmetry algebra of the relativistic theory reduces to corresponding non-relativistic symmetry algebra upon light-cone reduction, the QFT in this light-cone frame  evolves in light-cone time $x^+$ for fixed light-cone momentum $P_-$ and thus we obtain a system in $d + 1$ dimensions with non-relativistic invariance. This is known as discrete light-cone quantization of quantum field theories.

Since hydrodynamics is low energy fluctuation of equilibrium quantum field theory, light-cone reduction of relativistic constitutive equations boil down to the non-relativistic constitutive equations for a fluid in one lower dimension. We consider a relativistic anomalous fluid system in $(3+1)$ dimensions and reduce the constitutive equation over light-cone coordinates and obtain the corresponding non-relativistic equations for a fluid in one lower dimensions. We also find a mapping between the degrees of freedom of the $(d+2)$-dimensional fluid to the degrees of freedom of the $(d +1)$-dimensional fluid.

We denote the $d$ spatial coordinates with $x^i$. Metric components in light-cone coordinates are given by
\be \label{E:LCRgij}
g^{ij}=\delta^{ij} \quad\text{and}\quad g^{+-}=-1,
\ee
rest are zero. Gradient operator is given by,
\be \label{E:partialsDef}
\dow_{\mu}=\{ \partial_+,\partial_-,\partial_{i} \}
\quad\text{and}\quad \dow^{\mu}=\{
-\partial_-,-\partial_+,\partial_{i} \}. 
\ee
We shall reduce the theory along the $x^-$ direction, and consider $x^+$ to be the non-relativistic time. We consider only solutions to the relativistic equations that do not depend on $x^-$; that is, all derivatives $\pa_-$ vanish.

The reduction of relativistic conservation equations for energy-momentum and charge current are given by:
\begin{eqnarray}
\begin{split}
\partial_{+}T^{++}+\partial_{i}T^{+i}&=0, \quad
\partial_{+}T^{+-}+\partial_{i}T^{i-}=0\\
\partial_{+}T^{+j}+\partial_{i}T^{ij}&=0, \quad
\partial_+J^+ + \partial_i J^i=0.
\end{split}
\end{eqnarray} 
These equations are similar to non-relativistic equations 
\begin{eqnarray}
\begin{split}
\partial_t \rho+ \partial_i(\rho v^i)& =0,\quad
\partial_t (\rho v^i)+ \partial_i(t^{ij})=0,\\
\partial_t (\epsilon+ \frac{1}{2} \rho v^2)+ \partial_i(j^i) &=0,\quad
\partial_t Q+ \partial_i(j^i_q)=0
\end{split}
\end{eqnarray}
under following identifications
\begin{eqnarray} \label{Eq:identification}
\begin{split}
	T^{++}	& = \rho, \quad T^{i+}=\rho v^i,  \quad
	T^{+-}	= \varepsilon + \frac{1}{2}\rho \mathbf{v}^2, \quad
	T^{i-} =\varepsilon^i, \quad T^{ij} = t^{ij}, \\ 
	& \hspace{3cm}J^{+}	= Q, \qquad  J^{i}=j^i_q.
\end{split}
\end{eqnarray} 

We use these mappings to find the following relations between relativistic and non-relativistic parameters.
\begin{eqnarray}\label{eq:LCRdictionary}
\begin{split}
\rho	&= (E+P)(u^+)^2,\quad
v^i = \frac{u^i}{u^+} -\frac{\eta}{\rho}\mathbf{Y}^i, \quad p = P ,
\quad \varepsilon =\frac{1}{2}(E-P)\\
\varepsilon^i	&=  \left(\epsilon + p +\frac{1}{2}\rho \mathbf{v}^2\right)v^i
-\eta \sigma^{ik} v_k -\kappa_T \dow^i \t - \t \s \dow^i\left(\frac{\mu}{\t}\right)
\end{split}
\end{eqnarray} 
and
\begin{eqnarray}
\begin{split}
Q& =u^+n + u^+ \vp, \\
j^i &= Q v^i - \kappa_q \dow^i \t 
- \t \sigma_q \dow^i\lb\frac{\mu}{\t}\rb
-\tilde m\dow^i p
+ \tilde \k_q \e^{ij}\dow_j \t
+ \t \tilde{\sigma}_q\e^{ij} \dow_j\lb\frac{\mu}{\t}\rb
-\bar m\e^{ij} \dow_j p
\end{split}
\end{eqnarray} 
where,
\begin{eqnarray}\label{eq:diffdeffornrflu}
\begin{split}
\mathbf{Y}^\mu &=\left( \dow^\mu u^+ - \frac{u^+\dow^\mu P}{E+P} \right),\quad
\vp  =-\varrho u^\nu\dow_\nu \left(\frac{M}{T}\right) - \gamma u^\nu\dow_\nu T
- \mho \epsilon^{ij}u^+\nabla_i v_j.
\end{split}
\end{eqnarray}
Non-relativistic temperature, $U(1)$ chemical potential and charge density are given by
\be \label{E:tIden}
\t =\frac{T}{u^+}, \qquad \mu = \frac{M}{u^+}, \qquad Q =
n u^+ .
\ee
Non-relativistic transport coefficients are given by
\ben\label{eq:LCRdictionarytransport}
\begin{split}
	\eta &= \eta_{rel} u^+, \quad \kappa_T = \frac{\eta}{\t u^+}, \quad \s = \eta_{rel} \frac{Q}{\rho}, \quad \k_q = \frac{\k Q}{2(\varepsilon + p)}, \quad \t \s_q = \left[\varrho + \frac{\t \s Q}{2(\epsilon + p)} - \tilde m Q \t\right],\\
	\tilde m &= \frac{\gamma \t u^+}{2(\varepsilon + p)}, \quad \tilde\k_q = \k_T\frac{\mho (u^+)^2}{\eta}, \quad \tilde\s_q = \s \frac{\mho (u^+)^2}{\eta}, \quad \bar m = \frac{2 \mho (u^+)^2 }{\rho}.
\end{split}
\een
The above relations are the dictionary between relativistic and non-relativistic fluid data. Using fluid-gravity correspondence one can compute constitutive relations and transports for relativistic fluid \cite{Banerjee:2008th,Erdmenger:2008rm} and then using the above dictionary find transports for non-relativistic fluid \cite{Rangamani:2008gi,Brattan:2010bw}.

One can also check that starting from relativistic thermodynamic relations : $dE= TdS + M dn$ (first law) and $E+P = S T + n M$ (Euler relation) one obtains usual non-relativistic thermodynamics : $d\varepsilon=\tau ds+\mu dQ+\varrho_m d\rho$ (first law) and $\varepsilon+p = s\tau + \varrho_m \rho + \mu Q$ (Euler equation) if $\varepsilon+p+\varrho_m \rho =0$. The last equation puts a further restriction on non-relativistic thermodynamics.

%% file: charged.tex
\section{The bulk geometry for non-relativistic charged fluid} \label{sec:bulk}

In this section, we construct the bulk geometry that is dual to a non-relativistic conformal charged fluid theory living on its boundary. We begin with a locally boosted charged black brane solution in $4 + 1$ dimensional AdS space \cite{Banerjee:2008th,Erdmenger:2008rm}. We uplift the solution to 10 dimensions and then perform TsT transformations to obtain a new solution with a Schr\"odinger symmetry at the boundary. The effective five dimensional theory contains a metric, a mass-less gauge field, a massive gauge field and a dilaton. The mass-less gauge field corresponds to a global $U(1)$ current on the boundary.

\subsection{Locally boosted charged black brane geometry and
  hydrodynamics} 

The Einstein-Maxwell action with a gauge Chern-Simons term in $AdS_5$ space is given by,
\ben S= {1\over 16 \pi G_5} \int d^5x \sqrt{-g} \lb R + 12
-F_{ab}F^{ab} -\frac{4\k} 3 \e^{abcde}\cC_a\cG_{bc}\cG_{de}\rb .  
\een
The $AdS$ radius has been set to 1. Here, $\cC$ is an abelian gauge field
and 
\be
\label{eq:Gdef}
\cG_{ab} = \dow_a\cC_b-\dow_b\cC_a
\ee
is the corresponding field strength. The value of the parameter $\k$ is fixed : $\k=1/2\sqrt{3}$. However, we shall keep this parameter as $\k$ to trace the effect of this term in final results. Equations of motion obtained from this action are given by,
\ben
\begin{split}
  R_{ab} -\frac12 R g_{ab} - 6 g_{ab} -2 \lb \cG_{ac} \cG_b^{c}
  -\frac14 \cG^2 g_{ab}\rb &=0\\
  \nabla_b\cG^{ab}+ \k \e^{abcde}\cG_{bc}\cG_{de} &= 0.
\end{split}
\een
These equations of motion admit a solution for metric and gauge field, which in the boosted frame is given by (AdS-Reissner-Nordstrom black-brane solution)
\begin{eqnarray}
\label{eq:d3metricboost}
\begin{split}
ds^2 &=-2 u_\mu dx^\mu dr-r^2 f(r, q, m)
u_\mu u_\nu dx^\mu dx^\nu +r^2 (u_\mu u_\nu +\eta _{\mu\nu})dx^\mu
dx^\nu,  \\
\cC &=-\frac{\sqrt{3} q}{2 r^2} u_\m dx^\m
\end{split}
\end{eqnarray}	
with,
\begin{eqnarray}
f(r,m,q)=1-\frac{m}{r^4}+\frac{q^2}{r^6}.
\end{eqnarray}
The Hawking temperature, chemical potential and physical charge
density of this black brane are given by
\begin{eqnarray}
  T=\frac{1}{2 \pi}(2-q^2), \quad \phi= \sqrt{3}q/2 \quad
  n=\frac{\sqrt{3}q}{4 \pi G_5}. 
\end{eqnarray}
We choose the black hole horizon to be at 1. This implies a relation between the charge parameter $q$ and the mass parameter $m$ : $q^2=m-1$. This system is dual to a field theory with $U(1)$ gauge invariance living on the boundary of $AdS$ spacetime. Thermodynamic properties of this field theory can be computed from thermodynamic properties of charged black hole in the bulk.

Low energy fluctuations of a quantum field theory in local thermal equilibrium is best understood in terms of hydrodynamics. Different transport coefficients capture the response of the system against external probes. Such hydrodynamic properties of a boundary fluid system can be consistently studied holographically. Low energy fluctuations in the boundary correspond to low energy fluctuations of bulk parameters (velocity, temperature etc.). One can consistently solve the bulk geometry in presence of these perturbations order by order in derivative expansion \cite{Bhattacharyya:2008jc,Banerjee:2008th} by solving bulk equations of motion and construct stress tensor and other currents of the boundary system in presence of these perturbations using holographic dictionary \cite{Balasubramanian:1999re}. The stress tensor and currents are also written in terms of derivative expansion of velocities, temperature etc. First and higher derivative terms capture hydrodynamic properties of the system.

We consider velocities and temperature to be local function of boundary coordinates and expand them up to first order in derivative expansion \cite{Banerjee:2008th}
\begin{eqnarray}
\begin{split}
m&=m_0+x^{\mu} \partial_{\mu} m^{(0)},\quad
q=q_0+x^{\mu} \partial_{\mu} q^{(0)},\quad
u_i=x^{\mu} \partial_{\mu} u_i^{(0)}.
\end{split}
\end{eqnarray}
Here $u_i$ is $i^{th}$ component of velocity. Once we consider fluid velocities and temperature are local functions the metric and gauge field do not solve equations of motion any more. We need to add more terms to metric and gauge field to satisfy Einstein equation in the bulk up to first order in derivative expansion. This whole procedure was explicitly carried out in the \cite{Banerjee:2008th}. Here we write the final result in a covariant form.
\begin{eqnarray}\label{eq:5dchargedmetricglob}
\begin{split}
  ds^2& =-2u_{\mu}dx^{\mu} dr-r^2 f(r,m,q)
  u_{\mu}u_{\nu}dx^{\mu}dx^{\nu}+r^2 P_{\mu\nu}dx^{\mu}dx^{\nu}\\
  & \qquad +2u_{\mu}dx^{\mu}r(u^{\lambda}\partial_{\lambda}u_{\nu}-
  \frac{\partial_{\lambda}u^{\lambda}}{3 
  }u_{\nu}) dx^{\nu}+2 r^2 F_2(r,m)\sigma_{\mu\nu}dx^{\mu}dx^{\nu}\\
  & \qquad-2 u_{\mu}dx^{\mu}(\frac{\sqrt{3}\kappa q^3}{ m r^4}l_{\nu}+6 q r^2
  P_{\nu}^{\lambda} D_{\lambda}q F_1(r,m)) dx^{\nu}\\ 
  \cC&=\lb\frac{\sqrt{3} q}{2 r^2} u_{\mu}+\frac{\sqrt{3}\kappa q^2}{2 m
    r^2}l_{\mu}-\frac{\sqrt{3}r^5}{2} P_{\nu}^{\lambda} D_{\lambda}q
  F_1^{1,0}(r,m)\rb dx^{\mu}
\end{split}
\end{eqnarray}
where,
\begin{eqnarray}
\begin{split}
P^{\nu\mu} &=\eta^{\nu\nu} +u^{\nu}u^{\mu}, \quad \s^{\mu\nu} = P^{\mu\a}P^{\nu\b}\dow_{(\a}u_{\b)} -\frac13P^{\mu\nu}\dow\cdot u\\
  P^\l_\m D_{\lambda}q  &= P^\l_\m \dow_\l q + 3 q u^\l \dow_l u_\m
  \quad \text{is weyl covariant derivative and} \quad l^{\mu}
  =\epsilon^{\nu\lambda\sigma\mu} u_{\nu}\partial_{\lambda}
  u_{\sigma}\\
  F_1(r,m)
  &=\frac{1}{3}(1-\frac{m}{r^4}+\frac{q^2}{r^6})\int_{r}^{\infty} dp
  \frac{1}{(1-\frac{m}{p^4}+\frac{q^2}{p^6})^2}(\frac{1}{p^8}-\frac{3}{4
    p^7}(1+\frac{1}{m}))\\
  F_2(r,m)&=\int_{r}^{\infty} dp
  \frac{p(p^2+p+1)}{(p+1)(p^4+p^2-m+1)}.
\end{split}
\end{eqnarray}

The above solution has Poincar\'e isometry at the boundary. Calculating stress energy tensor and $U(1)$ charge current holographically \cite{Balasubramanian:1999re} one finds that these conserved currents describing a relativistic fluid up to first order in derivative expansion. The stress-energy tensor and charge current are given by,
\ben
\begin{split}
	T_{\mu\nu} & =(E+P)u_{\mu}u_{\nu}+P \eta_{\mu\nu} - \eta_{rel} \sigma_{\mu\nu}\\
	J_{\mu} &= n u^{\mu} -\cD P^{\nu}_{\mu} {D}_{\nu} n +\mho l_{\mu}
\end{split}
\een
where, transport coefficients are given by
\be\label{eq:holographictransportrel}
E=3P = {3m\over 16\pi G_5}, \quad \eta={1\over 16\pi G_5},\quad n={\sqrt3 q\over 4\pi G_5},\quad \cD = {1+m \over 4m}, \quad \mho = {3\kappa q^2\over 16 \pi G_5 m}. 
\ee
The relation $E=3P$, which is the equation of state for relativistic fluid, implies that the relativistic fluid is conformal and hence, relativistic fluid has zero bulk viscosity.

In the next subsection we start with locally boosted five dimensional geometry (\ref{eq:5dchargedmetricglob}) and construct a new solution with asymptotic Schr\"odinger isometry by using a solution generating technique called $TsT$ transformation.

\subsection{TsT transformation of locally boosted charged black brane solution}
\label{sec:tst}

The TsT transformation is a solution generating technique in string theory \cite{Son:2008ye,Herzog:2008wg,Balasubramanian:2008dm,Adams:2008wt,Goldberger:2008vg,Maldacena:2008wh,Bobev:2009mw,Bobev:2009zf,Banerjee:2011jb,Imeroni:2009cs,Yamada:2008if,Kim:2010tf,Singh:2010rt}. Supergravity is a low energy solution of full string theory. A solution to supergravity equations of motion can be shown to be a solution to string theory. A symmetry transformation of this solution with respect to a supergravity symmetry does not give rise to a new solution in supergravity. However, if the transformation employed is a symmetry of full string theory then the transformed solution can be interpreted as a new solution of string theory. In case of the TsT transformations, T-dualities are performed along the isometry direction, whereas s-transformation mixes the dualised coordinate with another isometry direction. This transformation generically changes asymptotic of the resulting metric but is guaranteed to be a solution to the supergravity equations of motion.

Since TsT transformation is a symmetry of full string theory we first need to uplift the five dimensional metric (\ref{eq:5dchargedmetricglob}, \ref{eq:d3metricboost} ) to ten dimensions and embed in string theory solution \cite{Mazzucato:2008tr,Imeroni:2009cs,Banerjee:2011jb}. The full type IIB 10 dimensional solution is a obtained by a direct sum of five dimensional metric and a Sasaki-Einstein space and a five form field strength.

Denoting this five dimensional metric
\be
ds^2_5 = g_{ab} dx^a dx^b, \quad \text{where $a,b$ are five
	dimensional indices}
\ee
the ten dimensional metric is given by
\be
ds^2_{10} = g_{AB}dx^A dx^B = g_{ab} dx^a dx^b+ ds^2_{SE}
\ee
where the Sasaki-Einstein manifold here is a five sphere, which can be
written as fibration over $\rm{CP^2}$
\be
ds^2_{SE} = \lb d\psi + \cP -\frac2{\sqrt3} \cC\rb^2 + ds_{\rm{CP^2}}^2\, .
\ee
The one form $\cP$ is given by
\be \cP = \frac{1}{3}
\left(\text{d$\chi $}_1+\text{d$\chi $}_2\right)-\sin ^2\alpha
\left(\text{d$\chi $}_2 \sin ^2\beta+\text{d$\chi $}_1 \cos
^2\beta\right), 
\ee 
and
\ben
ds_{\rm{CP^2}}^2 = d\alpha^2 &+& \sin^2\alpha d\beta^2 
+ \sin^2 \alpha\cos^2\alpha(\cos^2\beta d\chi_1 + \sin^2\beta
d\chi_2)^2  \nn
&+&  \sin^2\alpha \sin^2 \beta \cos^2 \beta (d\chi_1-d\chi_2)^2\, .
\een
The ten dimensional geometry is supported by a five form field strength is given by
\be
\cF_5 = 2(1+*_{10})\lB \lb d \psi +P-\frac{2}{\sqrt{3}} \cC \rb \wedge J_2-\frac{1}{\sqrt{3}}*_5 \cG\rB \wedge J_2, \quad \text{where} \quad J_2 = \frac12 d\cP,
\ee
a two form field strength $\cG$ given by equation (\ref{eq:Gdef}) and a dilaton
\be
\Phi=0.
\ee
To obtain TsT transformed solution we first define two light-cone coordinates $\xp$ and $\xm$
\be\label{eq:lccoord}
\xp = \gamma(v+z) \quad \tand \quad \xm = \frac1{2\g} (v-z)
\ee
where $\g$ is twist parameter defined below. Thus we have two isometry direction $x^-$ and $\psi$.  The TsT transformations, then, correspond to performing T-duality along $\psi$
direction, followed by a shift along $x^-$, i.e., $x^-\ra x^- - \g\psi$. $\g$ is shift or twist parameter, we often set that to $1$. Finally we perform T-duality back along the $\psi$ direction. 

To perform a TsT transformation we write down the five dimensional solution (\ref{eq:5dchargedmetricglob}) in the following form 
\be
ds^2 = -2 u_{\mu}dx^\m dr + \cS_{\m\nu} dx^\m dx^\n, \quad \cC = \cC_\m dx^\m
\ee
where,
\begin{eqnarray}
\begin{split}
\cS_{\m\nu}&=-r^2 f u_{\mu}u_{\nu}dx^{\mu}dx^{\nu}+r^2 P_{\mu\nu}dx^{\mu}dx^{\nu}
+2u_{\mu}dx^{\mu}r(u^{\lambda}\partial_{\lambda}u_{\nu}-\frac{\partial_{\lambda}u^{\lambda}}{3 }u_{\nu})dx^{\nu}\\ 
&\quad +2 r^2 F_2(r,m)\sigma_{\mu\nu}dx^{\mu}dx^{\nu}
-2 u_{\mu}dx^{\mu}(\frac{\sqrt{3}\kappa q^3}{ m r^4}l_{\nu}+6 q r^2 P_{\nu}^{\lambda} D_{\lambda}q F_1(r,m)) dx^{\nu},\qquad\\
\cC_\m &= \frac{\sqrt{3} q}{2 r^2} u_{\mu}+\frac{\sqrt{3}\kappa q^2}{2 m
	r^2}l_{\mu}-\frac{\sqrt{3}r^5}{2} P_{\nu}^{\lambda} D_{\lambda}q
F_1^{1,0}(r,m).
\end{split}
\end{eqnarray}
Replacing $v$ and $z$ in terms of light cone coordinates $x^\pm$ this metric can be written as,
\be
ds^2_5 = A_1(dx^-+K_1)^2 + ds_4^2
\ee
where
\ben
\begin{split}
	A_1 & = \cS_{--}, \quad K_1 = \frac1{\cS_{--}} \lb \cS_{+-} dx^+ -u_- dr
	+\cS_{- x}dx +\cS_{-y}dy \rb \\
	ds_4^2 &= -2u_{\bar a} dx^{\bar a} dr +\cS_{\bar a\bar b} dx^{\bar a}dx^{\bar b} +\frac1{\cS_{--}} [ (2u^-
	\cS_{--} -2 u^+ \cS_{+-}) dx^+ dr  \\
	& \quad - (u_- dr - \cS_{-i} dx^i)^2 -
	(\cS_{+-}dx^+ +2 \cS_{+i} dx^i) \cS_{+-} dx^+ ].
\end{split}
\een
Here $x^{\bar a} = \{x^+, x, y\}$ and $x^i = \{x,y\}$. Therefore, the full
10-dimensional metric is given by, Then 10 dimensional metric, therefore, is given by,
\ben 
ds_{10}^2=A_1(dx^-+K_1)^2 + ds_4^2 + \lb d\psi + \cP -\frac2{\sqrt3} \cC \rb^2 +
ds_{\rm{CP^2}}^2.
\een
There is no NS-NS two form and dilaton to start with. Apart from that we have a five form field strength and a two form field strength in 10 dimensions given by equation (\ref{eq:tsttranchar}).

A TsT transformation of the above 10 dimensional solution will give rise to the following 10 dimensional solution\footnote{We denote all the $TsT$ transformed fields with ``hat''. The above solution is written string frame.}.
\ben\label{eq:tsttranchar}
\begin{split}
	d\hat s^2 &= \cM A_1 (dx^-+K_1)^2 +\cM \lb d\psi + \cP-\frac{2}{\sqrt{3}} \cC \rb^2 +	ds_8^2, \quad e^{2\hat \Phi} = \cM, \\
	\hat B_2 & = \cA \wedge \lb d \psi + \cP-\frac{2}{\sqrt{3}} \cC\rb,  
	\quad \hat \cF_3 = g\wedge d\cP, \quad g = \frac12 d\cC_-,\\
	\hat \cF_5 &= \cF + \hat B_2 \wedge \hat \cF_3.
\end{split}
\een
where,
\ben
\cM = (1+\g^2 A_1)^{-1} \quad \text{and} \quad  \cA = -\g \cM A_1 (dx^-+K_1)
\een
In addition to this there is a two form field strength $\cG$ which remains unaltered under TsT transformation.

\subsection{Effective five dimensional action and boundary terms}
\label{sec:boundaryterm}

The $TsT$ transformed ten dimensional fields (\ref{eq:tsttranchar}) can be consistently truncated over $S^5$ \cite{Maldacena:2008wh,Adams:2009dm}. Upon KK reduction on $S^5$, the five-dimensional effective solution includes a metric, a massive vector $\cA_M$ coming from reduction of $\hat B_2$ over $S^5$, a scalar and a massless vector $\cC$, which was present in the 10D metric before the TsT. One can also obtain an one form from $\hat \cF_3$. However, this one form is not an independent mode of excitation. It is completely fixed by the massive gauge field and the massless gauge field. Massless gauge field doesn't change with  $TsT$ transformation and  remains same as before.  The reduced five dimensional fields in
Einstein's frame are given by \cite{Brattan:2010bw},
\ben\label{eq:tstmetein}
\begin{split}
	d\hat s_E^2 &= e^{4\Phi\over 3} A_1 (dx^-+K_1)^2 + e^{-{2\Phi\over 3}}
	ds_4^2\\
	\hat \cA &= -\g \cM A_1 (dx^-+K_1),\quad \hat \cC = \cC_\m dx^\m,  \quad g = \frac12 d\cC_-\\
	e^{2\hat \Phi}& = \cM e^{2\Phi}, \quad \cM = (1+\g^2 A_1)^{-1}.
\end{split}
\een

These five dimensional fields are solutions of equations of motions
(up to first order in derivative expansion) obtained from an effective
five dimensional action \cite{Brattan:2010bw}
\begin{eqnarray}
\label{eq:5deffactioncharged}
S_5=\frac{1}{16\pi G_5}\int d^5x \sqrt{-g}&&\Big[{R}-\frac{4}{3}(\dow_a \Phi)(\dow^a \Phi)-\frac{1}{4}e^{-8\Phi/3}\cW_{ab}\cW^{ab}-4\cA_a \cA^a+16 e^{2 \Phi/3}-4 e^{8 \Phi/3}\nonumber\\
&&-\frac{1}{3}e^{4\Phi/3}\cG_{ab} \cG^{ab}-4 e^{2\Phi}g_{a}g^{a}-e^{-4 \Phi/3}\cA_{a}\cG_{bc}\cA^{a}\cG^{bc}\nonumber\\
&&-\frac{1}{2}e^{-2\Phi/3}\left(-\frac{2}{\sqrt{3}}\cG_{ab}-4 g_a \cA_b\right)\left(-\frac{2}{\sqrt{3}}\cG^{ab}-4 g^a \cA^b\right)\nonumber\\
&&+ \frac{4\k}{3} \cG_{ab}\cG_{cd}\cC_{e}\epsilon^{abcde}
\Big].
\end{eqnarray}

Unlike asymptotically $AdS$ spacetime the metric (\ref{eq:tstmetein}) has inhomogeneous asymptotic fall-off. Therefore, it is not very straight forward to apply Brown-York type analysis to obtain boundary stress tensor for this geometry. First of all, to obtain a well defined boundary stress tensor one needs to add a suitable boundary term $S_{b}$ (which includes counterterms also) to action (\ref{eq:5deffactioncharged}) such that on-shell variation of the total action is zero and boundary stress tensor is finite. The boundary action for uncharged case is given in \cite{Herzog:2008wg,Adams:2008wt}. In presence of $U(1)$ charge, one can add extra counterterms for finite answer \cite{Liu:2004it}. However, since the leading massless $U(1)$ gauge field $\cC$ and its fluctuations fall sufficiently faster as we approach asymptotic boundary at $r\ra \infty$, we do not need any extra counterterms for renormalization. One can also have a $\ln r$ divergence at boundary \cite{DHoker:2009ixq} but in our case (the fluctuations we are considering) such divergences do not appear. Therefore, the boundary counterterms terms are given by 
\begin{eqnarray}
	\begin{split}
		\label{eq:actionb}
		S_{ct} &=\frac{1}{16 \pi G_5}\int d^4\xi \sqrt{-h} \lB-6 +3 \Phi^2 +\cA_{\mu}\cA^{\mu}  \rB .
	\end{split}
\end{eqnarray}
There are other terms, may be added to this boundary action \cite{Herzog:2008wg}, however those terms will not play any role in calculating boundary stress tensor complex and current. Hence we do not write them here.

\section{Holographic non-relativistic charged fluid}
\label{sec:stresstensor}

Following the holographic renormalization programme \cite{Balasubramanian:1999re,deHaro:2000vlm} in the gauge-gravity duality one can construct stress tensor for relativistic field theories from a well-defined action principle by varying the action with respect to the boundary data. In case of TsT transformed bulk geometries the asymptotic metric is not conformal to a flat spacetime. Non-relativistic time coordinate and space coordinate have different $r$ dependences, leads to degenerate boundary metric. It has been discussed in \cite{Ross:2009ar} how to obtain a consistent stress tensor complex for a non-relativistic system in presence of non-uniform $r$ dependence in boundary metric.

TsT transformed five dimensional bulk action (\ref{eq:5deffactioncharged}) includes a vector field  a scalar field and massless gauge field. These fields will modify the definition of holographic stress tensor and current. We follow the prescription of \cite{Hollands:2005ya} and \cite{Ross:2009ar} (who extended the work of \cite{Hollands:2005ya} to non-relativistic spacetime) to determine boundary stress energy complex for charged Schr\"odinger field theory. On-shell variation of bulk action and boundary counterterm action with respect to boundary fields is given by
\be
\delta S_5 +\delta S_{ct} = \frac1{16\pi G_5} \int d^4x\sqrt{-h} \lb s_{\a\b}\delta h^{\a\b}
+ s_{\a} \d \cA^\a +\tilde{s}_{\beta} \delta  \cC^{\beta}+ s_{\Phi}\d \Phi\rb,
\ee
where,
\begin{eqnarray}
\begin{split}
s_{\alpha \beta}&=\pi_{\alpha\beta}+3
h_{\alpha\beta}+\cA_\alpha
\cA_\beta-\frac{1}{2} \cA_\gamma \cA^\gamma
h_{\alpha\beta}-\frac{3}{2}\Phi ^2
h_{\alpha\beta}\\
s_\alpha&=-n^\mu \cW_{\mu\alpha} e ^{-8 \Phi /3}+2\cA_\alpha, \quad
s_\Phi =-\frac{8}{3} n^\mu \partial_\mu \Phi + 3\Phi.
\end{split}
\end{eqnarray}
Coefficient $\tilde s_\a$ is given by,
\begin{eqnarray}
\begin{split}
\tilde{s}_{\alpha}  &= -\lb \frac4{3}e^{\frac{4\Phi}{3}} + \frac{8}{3} e^{\frac{-2\Phi}{3}} + 4e^{-\frac{4\Phi}{3}} \cA_{b} \cA^{b} \rb n^{a}\cG_{a \alpha}
-\frac{8}{\sqrt{3}} e^{-\frac{2\Phi}{3}} n^{a} \lb g_{a} \cA_{\alpha} - g_{\alpha} \cA_a \rb\\
&\hspace{7.75cm} + \frac{8 \kappa}{3\sqrt{-h} }n^{a} \epsilon_{a\alpha b c d} \cC^{b} \cG^{cd}.
\end{split} 
\end{eqnarray}

The main idea of \cite{Ross:2009ar} is that in presence of tensor fields, we have to consider the variation of boundary frame field $\hat{e}_\alpha^{(A)}$ (instead of boundary metric), holding the tangent space components $\phi^{[\cI]}_{AB....}$ of the matter fields fixed where $A,B,..$ denote the tangent space indices and $\cI$ are for matter fields species. This treatment will provide a stress tensor which will give us conserved charges using the definition of charges from stress tensor, which will generate the symmetry of
asymptotic spacetime. This charges are conserved up to terms derivatives of other fields.

Writing metric in terms of frame fields and vector fields in terms of tangent space indices we have
\be
h^{\a\b} = e^\a_A e^\b_B \eta^{AB}, \quad \cA^\a = e^\a_A \cA^A, \quad \cC^\a = e^\a_A \cC^A.
\ee

Following the analysis similar to \cite{Ross:2009ar} we define components of boundary  charged non-relativistic stress energy complex and current as,
\ben
\begin{split}
	\delta S_b = \int d^4x\sqrt{-h} \left(T^{\alpha}_\b e^\b_A \delta
	{e}_{\alpha}^{A} + s_\a e^\a_A \delta A^A + s_{\phi} \delta \phi + J_\a e^\a_A \delta \cC^A\right).
\end{split}
\een
where,
\begin{eqnarray}
	\label{eq:nonstress1}
	\begin{split}
\varepsilon &= T^+_{\ +}=2 s^+_{\ +} - s^+ \cA_+-\tilde{s}^+ \cC_+,\hspace{1.1cm}
\varepsilon ^i = T^i_{\ +}=  2 s^i_{\ +}-s^i \cA_+-\tilde{s}^i \cC_+,\\
\rho^i &=  - T^i_{\ -}= - 2 s^i_{\ -}+s^i \cA_- + \tilde{s}^i \cC_-,\hspace{.7cm}
\pi^i_j  = - T^i_{\ j}=-2 s^i_{\ j}+s^i \cA_j+\tilde{s}^i \cC_j,\\
\rho &= T^+_{\ -}=2 s^+_{\ -}-s^+ \cA_-  -  \tilde{s}^+ \cC_-,\quad
\hspace{.75cm}Q = J^+ = \tilde{s}^+ ,\quad	j_i = J^i =  \tilde{s}^i.
	\end{split}
\end{eqnarray}
Thus addition of massless $U(1)$ field simply adds up a contribution to energy momentum complex like a massive gauge field. At the same time sources a global $U(1)$ charge current at the boundary : $J^0 $ and $J^i$ are charge density and charge current respectively.

Now we have all the terms required to do the computation of various components of stress energy tensor and current.

\subsection{Calculation of constitutive relations}

Writing mass and charge parameters ($m$ and $q$ respectively) of black hole in terms of temperature and chemical potential \cite{Son:2009tf}
\begin{eqnarray}\label{eq:mqrel}
\begin{split}
m&=\frac{\pi ^4T^4}{16}  (\gamma +1)^3 (3 \gamma -1), \quad
q=\frac{\phi }{\sqrt{3}}\frac{\pi^2  T^2}{2} (\gamma +1)^2\\
\text{where} \quad \g &= \sqrt{1+{8\phi^2\over 3\pi^2 T^2}}.
\end{split}
\end{eqnarray}
Here $\phi$ is chemical potential and horizon radius $r_+$ is given by
\be
r_+ = \frac{\pi}{2} T (\g+1).
\ee
Since, in our holographic calculations, unperturbed horizon radius is set to $r_+=1$, the unperturbed values of the parameters satisfy
\ben
m_0=1+q_0^2, \quad \phi_0 = \frac{\sqrt{3} q_0}{2}, \quad T_0 = \frac{2-q_0^2}{2\pi}, \quad \gamma_0 = \frac{2+q_0^2}{2-q_0^2}.
\een
$m_0,\ q_0, T_0, \phi_0$ are unperturbed values of $m,\ q, T, \phi$ respectively.

Now we compute non-relativistic constitutive relations order by order in derivative expansion. We start with ideal ($zero^{th}$) order.

\subsubsection{Ideal order}

We use equation (\ref{eq:nonstress1}) to compute mass density, energy density, pressure and charge at ideal order. They are given by 
\ben\label{eq:idealorder}
\r &= 4m_0 (u^+)^2, \quad
p =m_0, \quad \varepsilon = m_0, \quad Q =4\sqrt{3} \ q_0u^+.
\een
Following \cite{Banerjee:2014mka}, we define non-relativistic temperature and $U(1)$ chemical potential as
\be
\tau = {T\over u^+}, \qquad \text{and} \qquad \m = {\phi \over u^+}.
\ee
Entropy density of non-relativistic system at ideal order is given by the entropy density of black hole
\be
s= \frac{r_+^2 u^+}{4G_5} = 4\pi u^+.
\ee
We find that these ideal order variables satisfy non-relativistic Euler's relation 
\be
\tau s = \varepsilon+p - \mu Q - \varrho_m \rho
\ee
where $\varrho_m$, mass chemical potential associated with mass density $\r$, is given by
\begin{eqnarray}
\label{eq:masschem}
\varrho_m= -\frac{1}{2 (u^+)^2}.
\end{eqnarray}
Also note that non-relativistic system has equation of state $\varepsilon-p=0$ at ideal order. This is because the parent relativistic theory was conformal and had equation of state $E=3P$. Using holographic results for relativistic fluid (equation \ref{eq:holographictransportrel}) we see our holographic computation matches with LCR computations  (\ref{eq:LCRdictionary}) at ideal order : \(\varepsilon=\displaystyle\frac12(E-P)\) and \(p=P\). One can slao check that non-relativistic thermodynamic variables satisfy a constraint equation \(\varepsilon+p+\varrho_m\rho=0\) at ideal order.

\subsubsection{First order hydrodynamics}

We start with first order relativistic solution (\ref{eq:5dchargedmetricglob}) and write down the metric and gauge field in light-cone coordinates : $\{r, x^+, x^-, x, y\}$. We expand the velocity vector in derivative expansion in local rest frame and keep terms up to first order in derivative \cite{Banerjee:2008th}. We consider boundary indices ${\m,\n}$ running over 
 $\{+,-,x,y\}$. The velocity components are given by,
 \ben
 \begin{split}
 	u^+ & = 1 + \e \lb x^+ \dow_+ u^+ + x^j \dow_j u^+\rb, \quad
 	u^i = \e \lb x^+ \dow_+ u^i + x^j \dow_j u^i \rb.
 \end{split}
 \een
 $u^-$ component of velocity is fixed by normalization condition
 $u^2=-1$
 \be
 u^- =\frac12 -\frac\e2 \lb x^+ \dow_+ u^+ +x^i \dow_i u^i\rb
 +\cO(\e^2).
 \ee
 $\e$ is a derivative counting parameter. Mass and charge parameters are given as
 \ben
 \begin{split}
 	 m &= m_0+ \e(x^+\dow_+ m + x^i \dow_i m)\\
 	 q &=q_0+ \e( x^+\dow_+ q + x^i \dow_i q).
 \end{split}
 \een
 Writing the relativistic metric and gauge field in a local frame and in light-cone coordinates we perform TsT transformation as described in section \ref{sec:tst} to obtain five dimensional TsT transformed geometry (solution) as given by equation (\ref{eq:tstmetein}). The expressions are very lengthy, hence we refrain ourselves to produce those expression in the paper. As a consistency, one can check if these solutions satisfy effective five dimensional equations of motion obtained from the action (\ref{eq:5deffactioncharged}). We find that the TsT tranformed geoemtry satisfy 5D equations of motion provided 
 \begin{eqnarray}\label{eq:holoconstrain}
 \begin{split}
\partial_+u^+ & =\frac{1}{2} \partial_iu^i, \quad \partial_+m=- 4 m_0 \partial_+u^+, \quad \partial_i m=4 m_0 \partial _+ u^i ,\quad
 \partial_+q =-3 q_0 \partial_+u^+.
 \end{split}
 \end{eqnarray}
 These relation can also be obtained by doing the light cone reduction of relativistic constraint relations obtained in \cite{Banerjee:2008th}. 
 
Using the relations (\ref{eq:mqrel}), we express derivatives of mass and charge in terms of derivatives of temperature and chemical potential in the following way,
\begin{eqnarray}
\begin{split}
\dow_i m &= 4\pi \dow_i T + 4\sqrt3 q_0 \dow_i\phi \\
\text{and} \quad \dow_i q & = \frac{2 q_0 }{\gamma _0 T_0}\dow_iT + \frac{2(2+5q_0^2)}{\sqrt3(2+q_0^2)}\dow_i\phi.
\end{split}
\end{eqnarray}
Using the holographic constraints (\ref{eq:holoconstrain}) and equations in (\ref{eq:mqrel}) we find that
\be
\dow_+\lb \frac{\mu}{\tau}\rb =0
\ee
and 
\be
\dow_+ u^+ = -{1\over 2\tau} u^+\dow_+\tau.
\ee
In non-relativistic constitutive relations we have terms proportional to $\dow_iu^+$. We identify this term with derivative of mass chemical potential. From equation (\ref{eq:masschem}) we have,
\begin{eqnarray}
{\dow_iu^+ \over u^+} = -{\dow_i \varrho_{m} \over 2 \varrho_m}.
\end{eqnarray}
Thus, the constitutive relations for non-relativistic fluid can be written as derivative expansion of temperature, mass-chemical potential and $U(1)$ charge-chemical potential. However, we define a new basis to write the constitutive relations in a more compact form. We use a reduced chemical potential
\be
\dow_i\n \equiv \dow_i\lb \frac{\mu}{\tau}\rb 
\ee
such that $\dow_+\nu =0$ and a redefined mass chemical potential
\be
\dow_i \mu_m \equiv \dow_i u^+ =  u^{+3} \dow_i\varrho_m.
\ee
Hence, we display our constitutive relations as derivative expansion of $(\tau, \nu, \mu_m)$. From equation (\ref{eq:nonstress1}) we find\footnote{At first order, there will be some terms proportional to fluid velocity $v^i$. We have dropped those terms since we are working in a local frame.} non-relativistic stress tensor quantities energy density $\varepsilon$, energy current $\varepsilon^i$ , spatial stress tensor $\pi_{ij}$, mass density $\rho$ and mass current $\rho^i$. Here we have suppressed a factor of $16 \pi G_5$.
\begin{eqnarray}
\begin{split}
\r^i &= \r v^i, \quad
\pi^i_j = p \delta^i_j-\eta \sigma^i_j,\quad
\e_i = \lb \e+p+\frac12 \r v^2\rb v^i +\k_T \dow_i \t +\t \s \dow_i {\n} ,\\
j_i &= Q v_i +\k_q \dow_i \t +\t \s_q \dow_i {\n} +\l_q \dow_i \mu_m + \e^{ij} (\tilde \k_q \dow_i \t +\t \tilde\s_q \dow_i {\n} +\tilde\l_q \dow_i \m_m)
\end{split}
\end{eqnarray}
with,
\ben
\begin{split}
\r &= 4m_0 (u^+)^2,\quad
p =m_0,\quad
\varepsilon = m_0,\quad
v_i= {u_i\over u^+} +  \frac{1}{4 m_0 \tau _0} \dow_i \t +\frac{\nu_0 \tau _0^2 }{2m_0^2} \dow_i\n,  \\
Q & =J^0= 4 \sqrt{3} {q_0 u^+}+\Delta Q, \quad \Delta Q = - \frac{\l_q}{2\tau_0}\dow_+\tau+\frac{12 \kappa  q_0^2}{m_0}\epsilon ^{{ij}} \dow _i u_j .
\end{split}
\een
The transport coefficients in energy current are given by
\begin{eqnarray}
\label{eq:unchargedtransport}
\begin{split}
\eta &= u^+, \quad \kappa_T = -{1\over \tau_0}, \quad \s = -\frac{\sqrt3 q_0}{m_0} = -{Q u^+\over \rho}.
\end{split}
\end{eqnarray}
Transports appearing in charge current are given by,
\begin{eqnarray}
\begin{split}
\k_q &= \frac{\sqrt{3} q_0 \left(3 m_0+2\right)}{m_0 \tau _0},\qquad \sigma_q= \frac{\sqrt3 q_0 \tau_0 }{m_0}\kappa_q,\qquad \l_q = \frac{3\sqrt3 q_0(1+m_0)}{m_0}\\
\tilde{\kappa}_q &=- \frac{12 q_0^2 \kappa}{m_0 \ \tau_0}, \qquad \tilde{\sigma}_q =- \frac{12\sqrt3 \kappa q_0^3 }{m_0^2}, \qquad \tilde{\lambda}_q = -\frac{24 \kappa q_0^2}{m_0}.
\end{split}
\end{eqnarray}

Entropy density of the system receives a correction at first order. We denote that correction \cite{Banerjee:2014mka} by $\chi$
\be
s = 4\pi u^+(1+\chi).
\ee
Demanding that first law of thermodynamics will be satisfied at first order (in local frame), we have
\ben
\tau s u^+\lB \dow_+\chi + {\mu u^+\over \tau s} \dow_+ \Delta Q\rB =0.
\een
From this equation we find first order correction to entropy density
\ben
s =4\pi u^+ \lb 1- {\nu\over 4\pi} \Delta Q \rb.
\een
One can check that with this correction to entropy, Euler relation is satisfied for 
\be
\label{eq:masschem1st}
\varrho_m= -\frac{1}{2 (u^+)^2},
\ee
which means mass chemical potential does not receive any correction at first order and hence \(\varepsilon+p+\varrho_m\rho=0\) is also satisfied at first order in derivative expansion. This observation is in agreement with \cite{Banerjee:2014mka}.

%% file: compare.tex
\section{Discussion}
\label{sec:discussion}

In this paper we derive the constitutive relations of a charged \sch fluid in holographic set up. We start with a five dimensional geometry whose low energy fluctuations correspond to a relativistic charged conformal fluid with a parity-odd term in global $U(1)$ current \cite{Banerjee:2008th,Erdmenger:2008rm}. Then we uplift the five dimensional solution to ten dimensions which fits with a particular configuration in type IIB string theory. Using TsT transformation we generate a new solution of type IIB theory with has \sch isometry at the boundary. Upon reduction over $S^5$, the lower dimensional geometry serves as holographic dual of a \sch fluid at the boundary. Following the work of \cite{Ross:2009ar}, with a little modification in presence of $U(1)$ charge, we compute boundary stress tensor complex which provides constitutive relations of a \sch fluid.

One can also obtain a \sch fluid by reducing the constitutive relations of a relativistic theory over light-cone. This direct approach produces set of constitutive relations for a lower dimensional \sch fluid in terms of all the data of its mother theory \cite{Rangamani:2008gi,Brattan:2010bw,Banerjee:2014mka}. Similarly reducing the first law of thermodynamics and Euler relation for relativistic fluid one can recover first law and Euler relation for non-relativistic fluid but at the cost of an extra relation between the thermodynamic variables : $\varepsilon+p+\varrho_m\rho =0$. This extra constraint implies that the fluid we obtain under LCR is not a generic \sch fluid rather a restricted class. See section \ref{sec:review}. A generic \sch fluid can have more terms in constitutive relations \cite{Banerjee:2015uta,Banerjee:2015hra}. Therefore, we find it interesting to obtain the \sch fluid holographically and check if holographic results produce anything different than LCR. However, it turns out that both the approaches result the same fluid. Thermodynamic variables, computed holographically, also satisfy an extra condition $\varepsilon+p+\varrho_m \rho =0$ up to first order in derivative expansion. Probably there is no surprise in our results, as TsT transformation also involves a reduction of the bulk geometry along a light-cone direction to another geometry with \sch asymptotic. An explicit check of results enriches our understanding. 

We start with a locally boosted baulk geometry which is dual to a relativistic charged fluid with scale invariance. Scale invariance implies that the relativistic fluid has equation of state $E-3P=0$ and zero bulk viscosity. LCR of this fluid produces a non-relativistic fluid with equation of state $\varepsilon-p=0$ and with zero bulk viscosity (incompressible flow). We see that holographic computations also produces the same result. For uncharged fluid, transport coefficients obtained holographically (equation \ref{eq:unchargedtransport}) exactly matches with transports obtained by LCR (\ref{eq:LCRdictionarytransport}).

In case of charged fluid there are some apparent differences between holographic results and LCR. This is because \cite{Banerjee:2014mka} used a different basis ($\tau, \nu$ and $p$) than what we considered in this paper. However, if we look at first order correction to physical charge density we see that $Q$ has a parity-odd correction \(\displaystyle Q = 4\sqrt3 {q_0 u^+}- \frac{\l_q}{2\tau_0}\dow_+\tau +  \frac{12\kappa q_0^2}{m_0}\e^{ij}\dow_jv_j\). This correction term is same as we get after LCR (equation \ref{eq:diffdeffornrflu}). One striking difference is that holographic construction restricts the fluid to have time independent reduced chemical potential $\nu$.

\paragraph{The extra constraint for non-relativistic system :} The extra constraint relation $\varepsilon+p+\varrho_m\rho =0$ for \sch fluid when reduced holographically or by LCR  is natural\footnote{We are thankful to Nabamita Banerjee and Akash Jain for a discussion on this issue.}. In $(d+1)$ dimensions, a conformal relativistic fluid has $d+1$ constitutive relations. The number of variables which appear in constitutive relations and thermodynamics is $d+4$ ($d$ velocities, temperature $T$, energy $E$, pressure $P$ and entropy $S$). Among these variables one can take fluid velocity $u^i$ and temperature T as fluid variables and write energy momentum tensor as derivative expansion of $u^i$ and $T$ by solving constitutive equation. To solve the system exactly we need to know other thermodynamic variables (energy, pressure and entropy) in terms of temperature. Therefore we need three more equations. For relativistic fluid these equations are the first law of thermodynamics, Euler relation and equation of state :
\ben
\begin{split}
dP -SdT &= 0 \ \text{(1st law)},\\
E+P-ST &=0 \ \text{(Euler relation)}, \\ 
E-3P &=0 \ \text{(for conformal fluid)} .
\end{split}
\een
Solving these equations for relativistic fluid we have $S\sim T^3$, $E\sim 3 T^4$ and $P\sim T^4$ (up to an overall constant factor). Now consider a non-relativistic fluid in $d-1$ space dimensions (since LCR of a $(d+1)$ dimensional relativistic theory gives a non-relativistic system in $d-1$ space dimensions). This system has $d+1$ constitute equations : one continuity equation, $d-1$ momentum conservation equations and one energy conservation equations and ariables are  : $d-1$ velocities, temperature $\tau$, mass chemical potential $\varrho_m$, energy density $\varepsilon$, pressure $p$, entropy $s$ and mass density $\rho$. Using constitutive equations one can write mass current, energy current and stress tensor as derivative expansion of velocities, temperature and mass chemical potential. Therefore we are left with four thermodynamic variables $\rho, \varepsilon, s$ and $p$. To solve the system we have three equations as before :
\ben
\begin{split}
	dp -s d\tau - \rho d\varrho_m &= 0 \ \text{(1st law)},\\
	\varepsilon+p-s\tau -\varrho_m \rho &=0 \ \text{(Euler relation)}
\end{split}
\een
and equation of state. If we consider the non-relativistic fluid obtained from a relativistic system, then equation of state is given by $\varepsilon-p=0$. With this equation one can solve the above two equations (first law and Euler equation) and find : \(\displaystyle \varepsilon = p \sim \tau^{c_1}\varrho_m^{2-c_1} \) with $c_1+c_2 = 2$. To make it consistent with LCR (or holographic) results : \(\varepsilon=P, \ \tau = {T\over u^+} \ \text{and}\ \varrho_m=-{1\over 2(u^+)^2}\), we have $c_1=4$ and $c_2 =-2$. These values of $c_1$ and $c_2$ can also be fixed, in stead, imposing and extra condition $\varepsilon+p+\varrho_m\rho=0$. Therefore, when we synthesis a non-relativistic fluid from an exactly solvable parent relativistic fluid we get that extra equation. 

As mentioned in introduction, a more general class of \sch fluid can be obtained by light cone reduction of a modified relativistic system called "null fluid", a relativistic fluid with an extra null isometry direction \cite{Banerjee:2015uta,Banerjee:2015hra,Jain:2015jla,Jain:2016rlz,Banerjee:2016qxf}. A reduction of this system in its symmetry broken phase is equivalent to the non-relativistic fluid without any extra constraint relation. Thus we have freedom to choose this extra condition. LCR provides that relation, null fluid construction leaves that to us. In that sense, LCR deals with a particular type of fluid, where as null fluid is more general. It would, therefore, be interesting to find a bulk dual for null fluid.

Thus we find that holographic computation of non-relativistic constitutive relations are in one-to-one correspondence with those obtained by LCR up to first order in derivative expansion.

